\documentclass{desyproc}

\usepackage{amsmath,amssymb}
\usepackage{epsfig,graphicx}
\usepackage{subfigure}
\usepackage{graphicx}
\usepackage{rotating}
\usepackage{cancel}
\usepackage{bm}

\def\bra{\langle}
\def\ket{\rangle}
\def\beq{\begin{equation}}
\def\eeq{\end{equation}}
\newcommand{\C}[1]{\mathcal{#1}}
\def\ov{\overline}

\begin{document}

\title{\vspace{-3cm}
\hfill{\small{DESY 09-228}}\\[2cm]
Light Hidden $U(1)$s from String Theory}

\author{{\slshape Mark Goodsell}\\[1ex]
DESY, Notketra{\ss}e 85, 22607 Hamburg, Germany}


\desyproc{DESY-PROC-2009-05}
\acronym{Patras 2009} 
\doi  

\maketitle

\begin{abstract}
The possible masses and kinetic mixings of hidden $U(1)$s in the LARGE volume scenario are discussed, including the generalisation of the compact manifold to a $K3$ fibration. 
\end{abstract}

\section{Introduction}

Many of the talks at PATRAS 2009 (for example that of A. Linder) described laboratory experiments capable of detecting light hidden $U(1)$s; others (e.g. J. Redondo) discussed astrophysical and cosmological searches. As reviewed by J. Conlon, string compactifications generically give additional hidden gauge sectors, in particular
hidden U(1)s. This contribution aims to review how hidden U(1)s arise in LARGE volume string compactifications  \cite{LARGEVolumes} and their likely masses and interactions with the visible sector particles \cite{Goodsell:2009xc}.

The LARGE volume scenario involves IIB string theory compactified on a Calabi-Yau manifold having volume $\C{V}$ of the form
$$
\C{V} = \tau_b^{3/2} - h(\tau_i) \qquad \mathrm{or} \qquad \C{V} = \tau_{b'}^{1/2} \tau_{b} - h(\tau_i),
$$
where $h$ is a function of $\tau_i$, the K\"ahler moduli of ``small'' cycles; and $\tau_b$ is the modulus corresponding to a large cycle. The first case corresponds to a ``swiss cheese'' manifold; the second is the generalisation to a K3 fibration where now $\tau_{b'}$ represents the $K3$ fibre modulus. 

One small cycle contributes a non-perturbative superpotential and this leads to the stabilisation of the K\"ahler moduli at a non-supersymmetric minimum, provided that the complex structure moduli have first been stabilised by three-form fluxes and that there are more complex structure moduli than K\"ahler moduli. The volume is stabilised at a large value; as high as $5 \times 10^{27}$ (in units of the string length) for TeV scale strings, $5 \times 10^{13}$ for an intermediate string scale $M_s \sim 10^{10}$ GeV, or $\sim 50$ for GUT scale strings. The standard model is realised upon $D7$-branes wrapping some of the small cycles. 

In this scenario there are three classes of candidates for light $U(1)$s. One such  class are from (closed) Ramond-Ramond strings \cite{RRs}, counted by the number of complex structure moduli. These may kinetically mix \cite{Earlykm} with the hypercharge, but they have no matter charged under them, and since the LARGE volume scenario involves compactification on a K\"ahler manifold they do not have any axionic couplings and are therefore massless. Therefore they can only be detected by production of their gauginos \cite{Ibarra:2008kn}. 

We shall instead focus upon the open string $U(1)$s supported on branes, which may have masses and charged matter. For these $U(1)$s wrapping a cycle $\tau_i$ the gauge coupling is given by $g^{-2}_i = \frac{\tau_i}{2\pi g_s} .$
For branes wrapping small cycles these give gauge couplings of the same order as the hypercharge, but if the brane wraps the large cycle $\tau_b$, then the gauge coupling will be hyperweak with $g_b^{-2} \sim \frac{\C{V}^{2/3}}{2\pi g_s}$. In the case of a $K3$ fibration this can be even smaller; if $\tau_{b'} \ll \tau_b$ then we can in principle approach $g_b^{-2} \sim \frac{\C{V}}{2\pi g_s}$ (although we require $\tau_{b'} \gg \tau_i$).

\section{Kinetic Mixing}

\begin{center}\begin{figure}
\centerline{
\includegraphics[width=10cm]{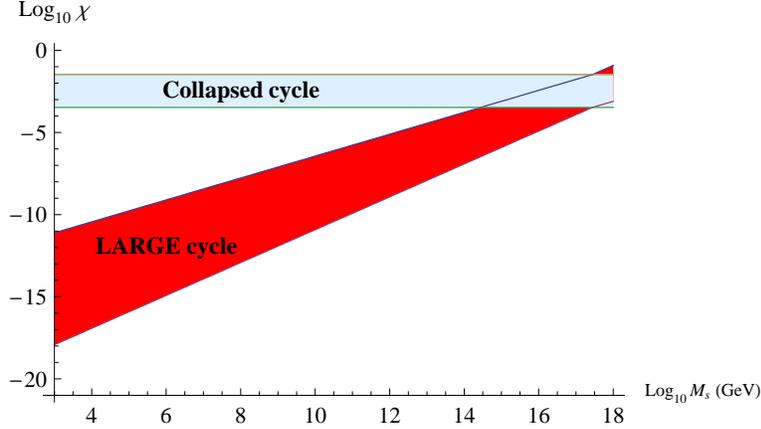}}
\caption{
Kinetic mixing between the visible U(1) and a U(1) sitting on a collapsed cycle (upper, blue) or
a hyperweak U(1) on a LARGE cycle (lower, red) as a function of the string scale.}
\label{Fig:chivsms}
\end{figure}\end{center}

If we assume that the additional $U(1)$s are hidden (in contrast to the $Z'$ scenario, see e.g. \cite{Langacker:2009su}), in that there is no light matter charged under both the visible and hidden sector fields, then we can only detect them via kinetic mixing with the hypercharge \cite{Earlykm}. The holomorphic kinetic mixing $\chi_{ab}^h$ between two gauge groups $a, b$ with holomorphic gauge couplings $g_a^h, g_b^h$, appears in the Lagrangian density 
$$\mathcal{L} \supset \int d^2 \theta  \left\{
\frac{1}{4 (g_a^h)^2} W_a W_a + \frac{1}{4(g_b^h)^2} W_b W_b
- \frac{1}{2}\chi_{ab}^h W_a W_b  \right\}, 
$$
and in type $IIB$ compactifications must have the form
$$
\chi_{ab}^h = \chi_{ab}^{\mathrm{1-loop}} (z^k, y_i) + \chi_{ab}^{\mathrm{non-perturbative}} (z^k, e^{-\tau_j}, y_i) ,
$$
where $z^k, y_i$ are the complex structure and brane position moduli respectively; the perturbative contributions cannot depend upon the K\"ahler moduli, and thus cannot be volume suppressed. After rescaling to the physical basis via the Kaplunovsky-Louis type relation \cite{Kaplunovsky:1994fg,Goodsell:2009xc}
$$
\frac{\chi_{ab}}{g_a g_b} = \mathrm{Re}(\chi_{ab}^h) + \frac{1}{8\pi^2} \mathrm{tr}\bigg( Q_a Q_b \log Z \bigg) -\frac{1}{16\pi^2} \sum_r n_r Q_a Q_b (r)\kappa^2 K,
$$
(where $K$ is the K\"ahler potential and $Z= \partial_{\alpha} \partial_{\ov{\beta}}K$ is the K\"ahler metric of matter fields) we find, since we are assuming no light matter charged under both hidden and visible sectors
$$
\chi_{ab} \sim \frac{g_a g_b}{16\pi^2} \ .
$$
This estimate is plotted in figure \ref{Fig:chivsms} for the case of branes on a collapsed (small, MSSM-like) cycle and on a LARGE cycle, taking into account the range of possibilities in the general K3 fibration scenario and allowing for an order of magnitude variation in the above estimate. 

There is also the possibility, should the kinetic mixing be cancelled, that it is generated by supersymmetry breaking effects; but in the LARGE volume scenario the values obtained are typically very small \cite{Goodsell:2009xc}.

\vspace{-1.2cm}
\section{$U(1)$ Masses}

Masses for $U(1)$s supported upon branes can be generated either via the St\"uckelberg mechanism or by explicit breaking with a charged field obtaining a vacuum expectation value. The latter could be due to a hidden Higgs mechanism or fermion condensate. We shall not discuss fermion condensates, as they would require some strong gauge dynamics in the hidden sector and the scale generated depends very sensitively upon the amount of hidden matter in the theory, so there is no generic prediction.

In the LARGE volume scenario anomalous $U(1)$s automatically obtain masses at the string scale, via the St\"uckelberg mechanism where the $U(1)$ is eaten by an axion. However, many non-anomalous $U(1)$s still obtain masses, but these generically contain some suppression by volume factors. There are two classes of axions that contribute; those counted by $h_-^{2,2}$ 
and those counted by $h_+^{1,1}$, respectively Hodge numbers odd and even under the orientifold. If we consider a simplified $2\times 2$ mass matrix of $U(1)$s where the first element corresponds to $U(1)$s on small (or collapsed) cycles and the second to one wrapping the LARGE cycle, then for the two types of contribution we have
$$
m^2_{{\rm St}\,(1)}
=  \frac{g_s}{2} M_s^2 \left( \begin{array}{ll}  \sim \mathcal{V}^{1/3} & \sim1 \\
\sim1 & \sim\mathcal{V}^{-1/3} \end{array} \right) , \qquad m^2_{{\rm St}\,(2)}
=  \frac{g_s}{2} M_s^2 \left( \begin{array}{ll}
\sim \mathcal{V}^{-1/3}  & \sim \mathcal{V}^{-2/3} \\
\sim \mathcal{V}^{-2/3} & \sim \mathcal{V}^{-1} \end{array} \right) .
$$
Thus if the a brane wraps a cycle that is anti-invariant under the orientifold projection then the first term will dominate. However, in early constructions of the LARGE volume scenario $h_-^{2,2}=0$. The second contribution arises only if the brane supports two-form fluxes. Thus a hyperweak gauge boson can acquire a mass $m_{\gamma'}$ as low as $\sim \mathrm{meV}$ if the string scale is $\sim \mathrm{TeV}$, for intermediate scale strings $m_{\gamma'} \sim \mathrm{TeV}$ but for a higher string scale the St\"uckelberg masses are beyond the reach of current experiments.

Finally turning to a hidden Higgs mechanism with hidden Higgs pairs $H_1, H_2$, the minimal potential is
$$
V = m_1^2 |H_1|^2 + m_2^2 |H_2|^2 + m_3^2 (H_1 H_2 + c.c) + \frac{1}{2}  (\xi_{\rm h} + g_{\rm h}|H_1|^2 - g_{\rm h}|H_2|^2)^2,
$$
where $m_1,m_2, m_3$ are soft masses and $\xi_{\rm h}=g_Y \chi_{ab} \frac{1}{8} v^2 \cos 2\beta$ is a Fayet-Iliopolous term generated by kinetic mixing with the hypercharge $D$-term, arising from the MSSM Higgs vev $v\simeq 246 \,\mathrm{GeV}$ and $\beta$ the angle parametrising the relationship between up and down Higgs vevs. If we take the hidden sector gauge coupling to be of the same order as the hypercharge and the soft masses to be generated by ``little gauge mediation'' from the visible sector, then the Fayet-Iliopoulos term generates a hidden gauge boson mass of $\sim \mathrm{GeV}$ \cite{DarkForces}. However, if we take the hidden gauge group to be hyperweak, then due to the very small kinetic mixing, we can generate in principle small masses since
$m_{\gamma'}^2 = 2g_{\rm h}^2  (|H_1|^2 + |H_2|^2)$.

If the symmetry breaking is dominated by the Fayet-Iliopoulos term, then  $m_{\gamma'} = 2 g_h \xi$ and the $m_i$ must necessarily be smaller than $g_{\rm h} \xi$, so that the Higgs mass is $\sim g_{\rm h} \xi \sim m_{\gamma'}$. Moreover, the above simple scenario leaves one Higgs field massless. This is a problem since the Higgs behaves like a minicharged particle, for which there are strict bounds if its mass is less than $\sim \mathrm{MeV}$. This problem persists if we set $m_i > \xi$ so that the hidden $U(1)$ is broken by an MSSM-type Higgs effect, since there $\bra H_1 \ket \sim \bra H_2 \ket \sim m_i/g_{\rm h} \rightarrow m_{\gamma'} \sim m_i$.

To obtain hidden photon masses smaller than $\sim \mathrm{MeV}$, there is a natural mechanism involving an additional hidden $U(1)''$ symmetry with coupling $\tilde{g}_{\rm h} \sim g_Y$ that obtains a mass $m_{\gamma''}$ via the St\"uckelberg mechanism. In this case, neglecting the Fayet-Ilioupoulos term, the potential is modified to 
$$
\tilde{V} = m_1^2 |H_1|^2 + m_2^2 |H_2|^2 + m_3^2 (H_1 H_2 + c.c) + \frac{1}{2} \bigg[g^2_{\rm h} + \tilde{g}^2_{\rm h} \left(\frac{  m_x^2}{ m_x^2 + m_{\gamma''}^2}\right)\bigg] (|H_1|^2 - |H_2|^2)^2 
$$
where $m_x$ is the mass of the modulus corresponding to the axion eaten by the $U(1)''$. We then obtain the relation
$$
m_{\gamma'} \gtrsim  \frac{1}{|W_0|} \ m_i \rightarrow m_{\gamma'} \gtrsim  \frac{1}{|W_0|} \ \mathrm{MeV},
$$
where $W_0$ is a constant parametrising the vacuum expectation value of the superpotential of the underlying supergravity theory. By taking this to be large we can obtain a hierarchy between the hidden gauge boson and Higgs masses, but at the expense of some fine-tuning.

\vspace{-0.25cm}
\section*{Acknowledgments}

I would like to thank my collaborators on this subject: Karim Benakli, Joerg Jaeckel, Javier Redondo and Andreas Ringwald; and for useful conversations Michele Cicoli, Joe Conlon, Nick Halmagyi, Amir Kashani-Poor, Sameer Murthy and Waldemar Schulgin.

\vspace{-0.25cm}
\begin{footnotesize}

\end{footnotesize}


\begin{thebibliography}{99}

\bibitem{LARGEVolumes}
V.~Balasubramanian, P.~Berglund, J.~P.~Conlon and F.~Quevedo,
JHEP {\bf 0503} (2005) 007;
C.~P.~Burgess, J.~P.~Conlon, L.~Y.~Hung, C.~H.~Kom, A.~Maharana and F.~Quevedo,
  JHEP {\bf 0807} (2008) 073;
M.~Cicoli, J.~P.~Conlon and F.~Quevedo,
  JHEP {\bf 0810} (2008) 105;
M.~Cicoli, C.~P.~Burgess and F.~Quevedo,
  JCAP {\bf 0903} (2009) 013;
J.~P.~Conlon, A.~Maharana and F.~Quevedo,
  JHEP {\bf 0905} (2009) 109.

\bibitem{Goodsell:2009xc}
  M.~Goodsell, J.~Jaeckel, J.~Redondo and A.~Ringwald,
  JHEP {\bf 0911} (2009) 027


\bibitem{RRs}
H.~Jockers and J.~Louis,
  Nucl.\ Phys.\  B {\bf 705} (2005) 167;
T.~W.~Grimm and A.~Klemm,
  JHEP {\bf 0810} (2008) 077;
T.~W.~Grimm, T.~W.~Ha, A.~Klemm and D.~Klevers,
  Nucl.\ Phys.\  B {\bf 816} (2009) 139;
A.~Arvanitaki, N.~Craig, S.~Dimopoulos, S.~Dubovsky and J.~March-Russell,
  arXiv:0909.5440.




\bibitem{Earlykm}
  J.~Polchinski and L.~Susskind,
  Phys.\ Rev.\  D {\bf 26} (1982) 3661;
B.~Holdom,
Phys.\ Lett.\  B {\bf 166} (1986) 196;
L.~B.~Okun,
Sov.\ Phys.\ JETP {\bf 56} (1982) 502
[Zh.\ Eksp.\ Teor.\ Fiz.\  {\bf 83} (1982) 892];
F.~del Aguila, G.~D.~Coughlan and M.~Quiros,
Nucl.\ Phys.\  B {\bf 307} (1988) 633
[Erratum-ibid.\  B {\bf 312} (1989) 751];
K.~R.~Dienes, C.~F.~Kolda and J.~March-Russell,
Nucl.\ Phys.\  B {\bf 492} (1997) 104
; S.~A.~Abel and B.~W.~Schofield,
Nucl.\ Phys.\  B {\bf 685} (2004) 150
; S.~A.~Abel, J.~Jaeckel, V.~V.~Khoze and A.~Ringwald,
Phys.\ Lett.\  B {\bf 666} (2008) 66 ;
S.~A.~Abel, M.~D.~Goodsell, J.~Jaeckel, V.~V.~Khoze and A.~Ringwald,
  JHEP {\bf 0807} (2008) 124 .

\bibitem{Ibarra:2008kn}
  A.~Ibarra, A.~Ringwald and C.~Weniger,
  JCAP {\bf 0901} (2009) 003.
\bibitem{Langacker:2009su}
  P.~Langacker,
  arXiv:0911.4294 [Unknown].

\bibitem{Kaplunovsky:1994fg}
V.~Kaplunovsky and J.~Louis,
Nucl.\ Phys.\  B {\bf 422} (1994) 57;
 K.~Benakli and M.~D.~Goodsell,
arXiv:0909.0017.



\bibitem{DarkForces}
N.~Arkani-Hamed and N.~Weiner,
JHEP {\bf 0812} (2008) 104;
E.~J.~Chun and J.~C.~Park,
JCAP {\bf 0902} (2009) 026;
C.~Cheung, J.~T.~Ruderman, L.~T.~Wang and I.~Yavin,
Phys.\ Rev.\  D {\bf 80} (2009) 035008;
D.~E.~Morrissey, D.~Poland and K.~M.~Zurek,
JHEP {\bf 0907} (2009) 050;
Y.~Cui, D.~E.~Morrissey, D.~Poland and L.~Randall,
JHEP {\bf 0905} (2009) 076;

















\end{thebibliography}
\end{document}